\documentclass{article}
\usepackage{graphicx}
\usepackage{amssymb}
\usepackage[note]{achemso}
\usepackage{jpc}

\def\C60{C$_{60}$}
\def\ch{C$_{70}$}
\def\cub{C$_8$H$_8$}
\def\ccm{C$_{60}\cdot$C$_8$H$_8$}
\def\ccp{poly(C$_{60}$C$_8$H$_8$)}
\def\hcm{C$_{70}\cdot$C$_8$H$_8$}
\def\hcp{poly(C$_{70}$C$_8$H$_8)$}
\def\cm-1{cm$^{-1}$}
\def\T1u{$T_{1u}$}
\def\Hg{$H_{g}$}
\def\Ag{$A_{g}$}

\begin{document}

\title{Vibrational spectra of C$_{60}\cdot$C$_8$H$_8$ and C$_{70}\cdot$C$_8$H$_8$ in the rotor-stator and polymer
phases}

\author{G. Klupp\thanks{Authors to whom correspondence should be addressed.
Email: klupp@szfki.hu (G. Klupp), kamaras@szfki.hu (K. Kamar\'as)} \thanks{Research Institute for Solid
State Physics and Optics, Hungarian Academy of Sciences} ,
F. Borondics$^{\dag}$, \'E. Kov\'ats$^{\dag}$, \'A. Pekker$^{\dag}$,\\
G. B\'enyei\thanks{Department of Organic Chemistry, E\"otv\"os
Lor\'and University}, I. Jalsovszky$^{\ddag}$,
R. Hackl\thanks{Walther Meissner Institute, Bavarian Academy of Sciences and Humanities} ,
S. Pekker$^{\dag}$, K. Kamar\'as$^{*\dag}$\\[12pt]
Research Institute for Solid State Physics and Optics,\\ Hungarian
Academy of Sciences,\\ P.O. Box 49,
H-1525 Budapest, Hungary;\\
Department of Organic Chemistry, E\"otv\"os Lor\'and University, \\Budapest, Hungary;\\
Walther Meissner Institute, Bavarian Academy of Sciences and Humanities, \\
85748 Garching, Federal Republic of Germany }

\maketitle

\newpage
\begin{abstract}

\ccm\ and \hcm\ are prototypes of rotor-stator cocrystals. We present
infrared and Raman spectra of these materials and show how the
rotor-stator nature is reflected in their vibrational properties. We
measured the vibrational spectra of the polymer phases \ccp\ and
\hcp\ resulting from a solid state reaction occurring on heating.
Based on the spectra we propose a connection pattern for the
fullerene in \ccp, where the symmetry of the C$_{60}$ molecule is D$_{2h}$.
On illuminating the \ccm\ cocrystal
with green or blue light a photochemical reaction was observed
leading to a similar product to that of the thermal polymerization.
\end{abstract}


\newpage
\section{Introduction}
\label{intro}

Fullerenes and cubane have recently been shown to form so called
rotor-stator cocrystals.\cite{pekker05} These cocrystals are
different from both orientationally ordered and plastic crystals, as
one of their constituents (the fullerene) is rotating and the other
one (the cubane) is fixed in a well-defined orientation. In the case
of \ccm\ rotating \C60\ molecules form a face centered cubic lattice
and static cubane molecules, occupying interstitial octahedral
sites, serve as bearings between them. \hcm\ crystallizes in a
face-centered cubic structure above 375~K.\cite{BortelKirch} At room
temperature the rotation of \ch\ is somewhat restricted, which leads
to a tetragonal distortion; the \ch\ molecule is able to rotate
around its main axis which, in turn, precesses around the
crystallographic $c$ axis. The formation of these structures is
driven by the molecular recognition between the concave surface of
the cubane and the round surface of the
fullerenes.\cite{pekker05,PekkerKirch}

On heating the fullerene-cubane compounds undergo a topochemical
reaction.\cite{pekker05} As the reaction product is insoluble in
common solvents, it is most likely a copolymer of the fullerene with
cubane.\cite{KovatsKirch} X-ray diffraction patterns of the annealed samples, measured at
room temperature, show a large emerging
amorphous part and weakening reflections compatible with fcc
structure. Compared to the original monomer phase the shift of these
reflections indicates lattice expansion and their intensity quickly
vanishes at high angles. Because of the parallel appearance of the
amorphous  contribution and disappearance of crystallinity we can
assume that the amorphous phase retains the local cubic order.
Another observation which makes this assumption reasonable is that
the morphology of the crystals does not change on
heating.\cite{pekker05}

In this paper we present a detailed vibrational (infrared and Raman)
characterization of the monomer and polymer phases of \ccm\ and
\hcm. In the monomer phases, we can confirm the rotor-stator nature
of the materials. Based on the spectra of the polymer phases, we
deduce the symmetry of the majority of the fullerene units as D$_{2h}$,
similar to the linear cycloaddition polymers. This conclusion is consistent with
a substantial presence of linear segments in the copolymer.

We published the infrared spectra of the monomer and polymer phases of
\ccm\ and \hcm\ earlier as supplementary material to Ref. 1. A
thorough study of polymerization of \ccp\ at high temperature and
high pressure has been performed by Iwasiewicz-Wabnig \emph{et
al.},\cite{iwasiewicz07} using x-ray diffraction and Raman
spectroscopy. Our results, obtained at ambient pressure on
annealing, are complementary to that study, except that we observe a
photopolymerization reaction on illumination with green or blue
light, which accounts for the laser wavelength dependence of the
Raman spectra.

\section{Experimental methods}
\label{exp}

Cubane was prepared following the method of Eaton and Cole.\cite{eaton2-64} Cubane and the fullerenes C$_{60}$
and C$_{70}$ were coprecipitated from toluene by adding isopropyl alcohol or by evaporating the solvent  to form
C$_{60}\cdot$C$_8$H$_8$ and C$_{70}\cdot$C$_8$H$_8$.\cite{pekker05}

The resulting black powder was pressed into KBr pellets for infrared
(IR) measurements. The spectra were recorded by a Bruker IFS28 and a
Bruker IFS 66v/S spectrometer. Depending on the width of the lines
to be observed, the resolution was set between 2 and 0.25 cm$^{-1}$.
Temperature-dependent measurements were conducted in a flow cryostat
cooled by liquid nitrogen or helium with the temperature adjustable
between 20 and 600~K. The KBr pellet technique has the disadvantage
that the index of refraction of the samples is generally in mismatch
with that of the medium, therefore the lineshapes become asymmetric
(Christiansen effect). However, the alternative of using organic
oils as Nujol was discarded because we wanted to identify as many
new infrared lines as possible, without disturbing absorption from
the medium.

Raman microscopy data were acquired in backscattering geometry on
powder samples either under ambient conditions or in an evacuated
glass capillary. Spectra were taken with three lines (468~nm, 531~nm
and 676~nm) of a Kr-ion laser on a triple monochromator (Jobin-Yvon
T64000). The laser power was carefully adjusted not to cause
polymerization or any other type of changes in the samples. This was
guaranteed with a power of 70-100~$\mu$W focused to a spot of
approximately 2~$\mu$m diameter. The slit width was set at 300 or
400~$\mu$m. For these small crystals (typically less than 10 $\mu
m$) the orientation of the principal axes with respect to the
polarization of the incident (${\bf e}_i$) and the scattered (${\bf
e}_s$) light could not be determined. However, in case of highly
symmetric molecules the fully symmetric $A_g$ vibrations can easily
be identified by comparing polarized (${\bf e}_s\parallel {\bf
e}_i$) and depolarized (${\bf e}_s \perp {\bf e}_i$) spectra. For
simplicity we label these by $xx$ and $xy$, respectively. The Raman
spectra taken with the 785~nm laser line of a diode laser were
collected by a Renishaw 1000 MB Raman spectrometer.

\section{Results and discussion}
\label{results}

\subsection{Rotor-stator phases}

\begin{figure}[p]
\includegraphics{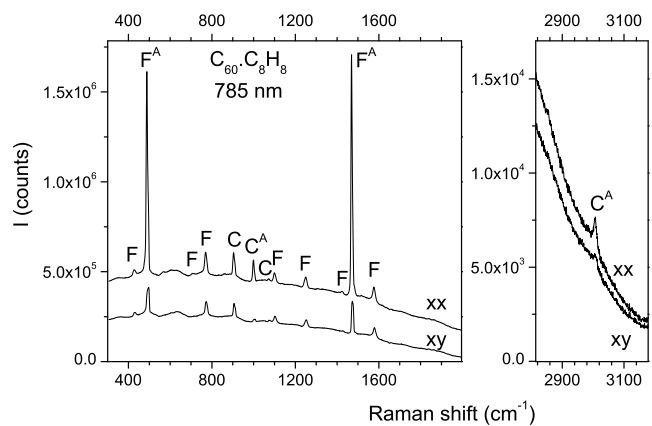}
\caption{Room temperature Raman spectra of the \ccm\ cocrystal.
The diode laser was operated at the line indicated. Spectra taken
with the incident and scattered light polarizations parallel and
perpendicular are labelled by $xx$ and $xy$, respectively. Cubane
modes\cite{della79} are denoted by C, fullerene modes
\cite{bethune91} by F. Totally symmetric modes are marked by
superscript A.} \label{fig:C60monRam}
\end{figure}

\begin{figure}[p]
\includegraphics{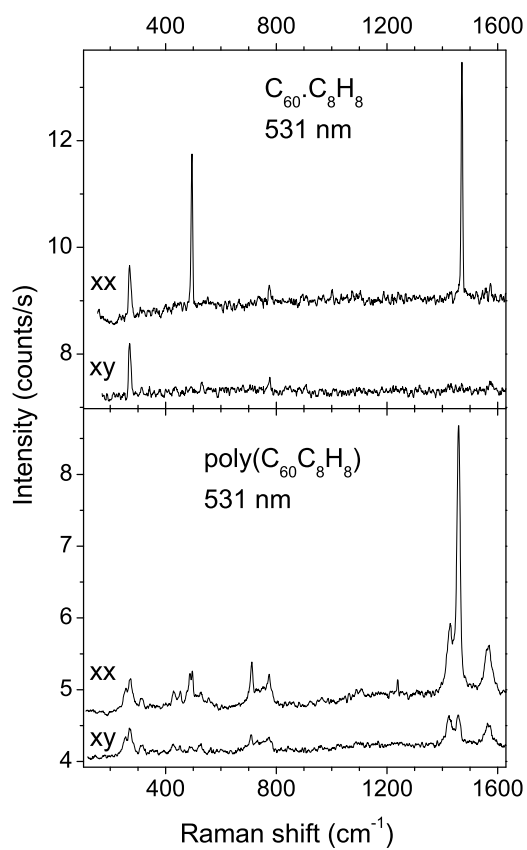}
\caption{Raman spectra of \ccm\ at room temperature before
annealing (monomer) and after annealing at 470~K (polymer). The
Kr$^+$ laser line and the polarizations are indicated. The spectra
are vertically shifted for clarity.} \label{fig:C60Ram}
\end{figure}

\begin{figure}[p]
\includegraphics{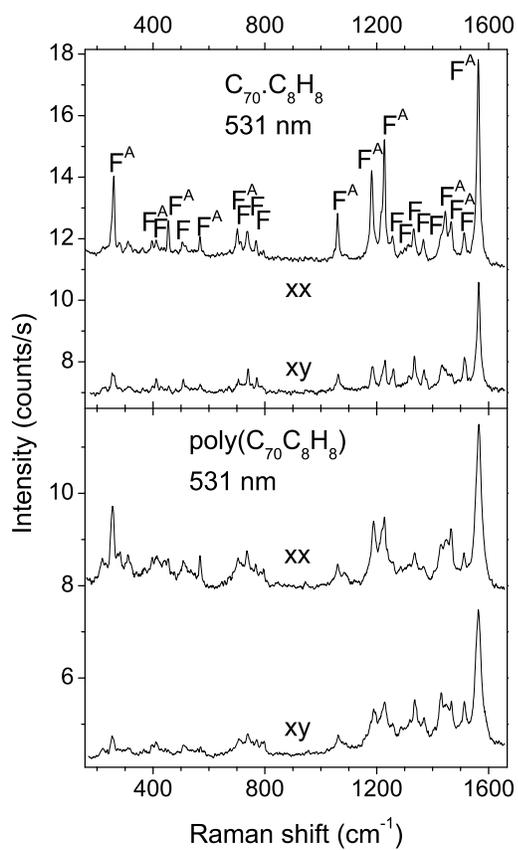}
\caption{Room temperature Raman spectra of \hcm\ cocrystal
and copolymer. The Kr$^+$ laser line and the polarizations are
indicated. The spectra are vertically shifted for clarity. Totally
symmetric modes are denoted by superscript A.\cite{sun02} Fullerene
peaks are marked by F, \cite{bethune91} no cubane peaks were found.}
\label{fig:C70Ram}
\end{figure}

\begin{figure}[p]
\includegraphics{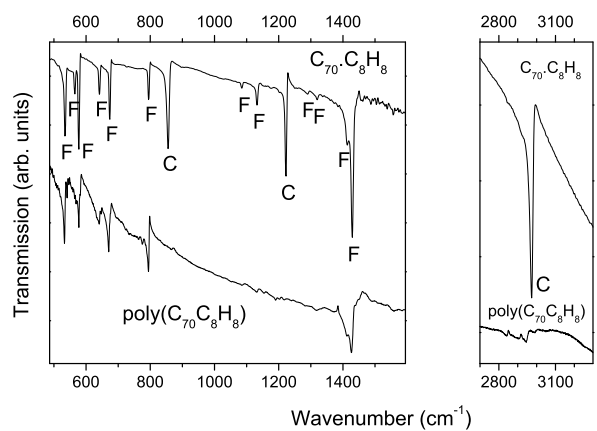}
\caption{Infrared spectra of \hcm\ before and after
annealing at 470~K (cocrystal and copolymer phase, respectively). C:
cubane peaks, \cite{della79} F: fullerene peaks \cite{bethune91}.
The asymmetric line shape is due to the Christiansen effect.}
\label{fig:cc70heating}
\end{figure}

The Raman and infrared spectra of \ccm\ in the rotor-stator phase
are shown in Figs. \ref{fig:C60monRam}, \ref{fig:C60Ram} and
\ref{fig:cc60heating} and those of \hcm\ in Figs. \ref{fig:C70Ram}
and \ref{fig:cc70heating}. The frequencies of the observed
vibrational peaks of \ccm\ are listed in Tables \ref{table:c60r} and
\ref{table:c60ir}, and those of \hcm\ in Tables \ref{table:c70r} and
\ref{table:c70ir}. We compare these frequencies to experimental data
on cubane\cite{della79} and C$_{60}$ (Ref. \citenum{bethune91}) and
calculated Raman\cite{sun02} and infrared\cite{stratman98} spectra
of C$_{70}$, respectively. As expected for molecular cocrystals with
the lattice stabilized by van der Waals interaction only, the
spectra are superpositions of those of the constituents. As no
crystal field splitting of the fullerene lines is observed, the
infrared measurement confirms that the fullerene molecules are
rotating in the crystal. The cubane lines are not split either,
proving that the crystal field around the cubane has the same point
group, i.e. $O_h$, as that of the isolated molecule.\cite{pekker05}
In the Raman spectrum of the rotor-stator cocrystals taken with 785~nm
excitation the fullerene lines are significantly stronger than the
cubane lines, most probably because of the enhanced Raman cross
section caused by the conjugated bonds, similarly to what was found
in fullerene clathrates.\cite{kamaras93} This effect renders the
cubane lines almost unnoticeable. When changing the wavelength of
the exciting laser to 531~nm, all of the cubane lines are lost
(Fig.~\ref{fig:C60Ram}), because we approach resonant scattering in
the fullerenes.\cite{matus91}

\C60 belongs to the icosahedral ($I_h$) point group and consequently shows four infrared-active vibrational modes
with \T1u\ symmetry. Out of its ten Raman-active modes, two belong to the \Ag\ and eight to the \Hg\ irreducible
representation. We could observe all of these modes in the spectrum of \ccm\ (the \Hg(1) mode can be seen in Fig.
\ref{fig:C60Ram}). \ch\ has $D_{5h}$ symmetry and altogether 31 IR active and 53 Raman active vibrational modes.
The IR modes can be decomposed as 21~$E'_1$ + 10~$A''_2$, and the Raman modes as 12~$A'_1$ + 22~$E'_2$ +
19~$E''_1$. Similarly to the case of pristine \ch, not all of these modes have sufficient intensity to be easily
detected. \cite{bethune91} Cubane belongs to the octahedral ($O_h$) point group. Its three infrared-active \T1u\
modes are clearly visible in the spectra of the \ccm\ and \hcm\ rotor-stator cocrystals. This cubane spectrum is
indeed the closest to that of isolated cubane in a crystalline environment; solid cubane\cite{della79} shows a
more complicated spectrum because of the lower site symmetry. The eight Raman-active modes of cubane are
classified as 2~$A_{1g}$ + 2~$E_{g}$ + 4~$T_{2g}$. Only three out of these eight appear in the \ccm\ spectrum
taken with the 785~nm laser and none in the spectra taken with the 531~nm laser, because of the aforementioned
cross-section differences.

In the \ccm\ cocrystal, the depolarization ratio $\rho =
\frac{\phi_{xy}}{\phi_{xx}}$ (with $\phi_{ij}$ the oscillator
strength of an excitation at either $xy$ or $xx$ polarization; see
section 2) should be zero for the fullerene \Ag\ modes and
$\frac{3}{4}$ for the \Hg\ modes. The \Ag\ modes were indeed found
totally polarized, and the depolarization ratio was 0.90 for the
\Hg(1) and 0.71 for the \Hg(4) mode (see Fig. \ref{fig:C60Ram}). In
contrast the totally symmetric modes of \ch\ should not vanish
completely in the $xy$ geometry because of its $D_{5h}$ symmetry.
This is what is found in the \hcm\ cocrystal. The modes that have
lower depolarization ratios are labeled by A in Fig.
\ref{fig:C70Ram}. These modes correspond to the ones assigned to
$A'_1$ by Sun and Kert\'esz. \cite{sun02}

In contrast to the fullerenes, the frequencies of the cubane principal lines in the rotor-stator crystals deviate
from those of cubane in its pure solid form. \cite{della79} If we compare the vibrational frequencies for various
environments of the cubane molecule, a clear trend can be observed. The highest vibrational frequencies occur in
the gas phase.\cite{cole81} In pure solid cubane or in solution the lines shift to lower frequencies. Further
downshift is found in \ccm\ and finally in \hcm. This trend is similar to that found in the vibrational
frequencies of molecules trapped in rare gas matrices \cite{abe03} and is caused by van der Waals interaction:
the higher the polarizability of the environment, the lower the frequency. The relatively large shifts in the
solids reflect the high polarizability of the fullerenes.

\subsection{Poly(C$_{60}$C$_8$H$_8$)}

\begin{figure}[p]
\includegraphics{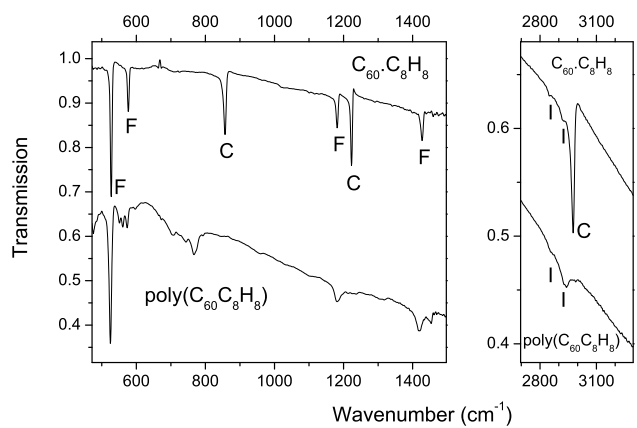}
\caption{Infrared spectra of \ccm\ before (cocrystal)
and after annealing at 470~K (copolymer). C stands for cubane
modes, \cite{della79} F for fullerene modes, \cite{bethune91} and I
for impurity. The spectra are vertically shifted for clarity. The
changes in the spectra show that annealing leads to the
polymerization of the sample.} \label{fig:cc60heating}
\end{figure}

The spectra of \ccm\ change dramatically upon annealing to 470~K
either in a furnace or in a heated cryostat in the IR spectrometer
(Fig. \ref{fig:cc60heating}). The Raman and IR spectra of the
annealed sample are plotted in Figs. \ref{fig:C60Ram} and
\ref{fig:cc60heating}, and the peak positions listed in Table
\ref{table:c60r} and \ref{table:c60ir}, respectively. Upon heating
to 470~K an irreversible reaction takes place. When annealing a few
tens of mg sample in the furnace, the first changes in the IR
spectra appear after 40 minutes: \C60\ modes split and new modes
appear. Further annealing leads to the disappearance of the original
\C60\ and cubane modes and increased intensity of the new peaks. The
new features of the final reaction product in the IR spectrum are
the same, irrespective of whether the annealing was done in a
furnace or \textit{in situ} in a cryostat.

In the Raman spectrum of the annealed \ccm\ the \Ag\ modes of \C60
do not split, but the low energy, i.e. radial \Hg\ modes show at
least a threefold splitting, best seen on the lone-standing \Hg(1)
mode. In the IR spectrum the original \T1u modes of the fullerene
split into at least two lines, and new peaks appear between 700
and 800 \cm-1. The splitting and the new modes indicate that the
\C60\ molecule is distorted. However, the number of new lines is
considerably less than would be expected if the cage
opened.\cite{vougioukalakis04} In contrast, the change in the
cubane lines is striking. The original lines disappear completely,
only a weak IR line at 2948 \cm-1\ indicates that there are still
hydrocarbon groups in the sample. We infer from the position of
this line, which corresponds to the C-H stretching in saturated
hydrocarbons, that the carbon atoms involved are sp$^3$
hybridized. In the reaction, we have to account for all atoms
since no mass loss was observed by thermogravimetry-mass
spectrometry (TG-MS) up to 570 K.\cite{pekker05} This suggests
that the cubane transforms into a different constitutional isomer
and covalently bonds to \C60, leading to a structural distortion.
The reaction product is most probably a covalently bound
copolymer, as the products are insoluble in common solvents.

Pristine cubane also isomerizes at 470~K,\cite{hassenruck89} the
same temperature where the polymerization appears in \ccm. Hence, a
straightforward assumption is that the first step of the
copolymerization reaction must be the decomposition of cubane.
Pristine cubane can decompose into several products, e.g.
cyclooctatetraene, bicyclooctatriene, styrene and
dihydropentalene.\cite{hassenruck89} As the first three form known
adducts with \C60,\cite{ishida00} which we could not detect by
either IR spectroscopy or HPLC\cite{KovatsKirch}, we can exclude
these as being the connecting units between the fullerenes.

In principle both fullerene-fullerene and fullerene-\cub\ bonds can
be realized in the polymer. \cub-\cub\ bonds can be excluded, as the
C$_8$H$_8$ molecules are well separated by the fullerene molecules.
We can also exclude the possibility of covalent fullerene-fullerene
bonding because of the following experimental observations. There
are two known bond types between fullerene molecules in fullerene
homopolymers. In neutral polymers the [2+2] cycloaddition leads to a
cyclobutane-type ring with two single bonds between the buckyballs.
\cite{zhou93,nr95} A Raman peak at approximately 950 \cm-1\ is
associated with this bond. \cite{wagberg99} This peak is absent in
the spectrum of \ccp. The other possible bond type is
one single bond between two fullerene molecules. \cite{oszlanyi96}
This bond leads to the appearance of a characteristic IR peak
between 800-850 \cm-1. As this peak is also absent we can rule out
the fullerene-fullerene direct bond. There is still another
observation which confirms this assumption. In fullerene polymers
\cite{raosci,lebedkin98} and in the dimer-oxide C$_{120}$O
\cite{krause98,lebedkin98} interball vibrational peaks appear in the
Raman spectrum between 100-140 \cm-1. We measured the Raman
spectrum down to 20 \cm-1, but did not find any peaks below the
split \Hg(1) mode. The reason for the absence of the interfullerene
bonding comes from structural considerations. The large
interfullerene distance observed by x-ray diffraction
\cite{pekker05} does not allow the \C60\ molecules to approach each
other close enough for a reaction to occur between them.

In the following we try to establish the connection pattern of the
fullerene unit based on the infrared and Raman spectra. Since the IR
and Raman spectra retain mutual exclusion (no lines are observed to
appear simultaneously in both), the inversion center of the C$_{60}$
balls must be preserved. This means that the possible point groups
of the \C60\ molecules are: $I_h$, $T_h$, $S_6$, $D_{5d}$, $D_{3d}$,
$D_{2h}$, $C_{2h}$ or $C_i$. In Table~\ref{table:correl} we show the
evolution and splitting of the Raman active \Ag\ and \Hg\ and the IR
active \T1u\ modes caused by symmetry reduction from I$_h$ to these
point groups (correlation table). The $C_{2h}$ and $C_i$ point
groups can be ruled out because the expected number of additionally
activated peaks\cite{Kirch03,long07} is too high to be reconciled
with the observed data. A $D_{2h}$ distortion could in principle be
positively identified as it leads to a threefold splitting of the
\T1u modes, in contrast to the others; unfortunately, in this case
our fits were not sufficiently robust to distinguish between a
three- or twofold splitting. $I_h$ or $T_h$ symmetry would not cause
splittings, therefore these cannot be the only point groups
appearing; there must be units of reduced symmetry even if the
connection pattern of the fullerene units is not uniform throughout the
whole polymer.

\begin{figure*}[p]
\includegraphics[scale=0.2]{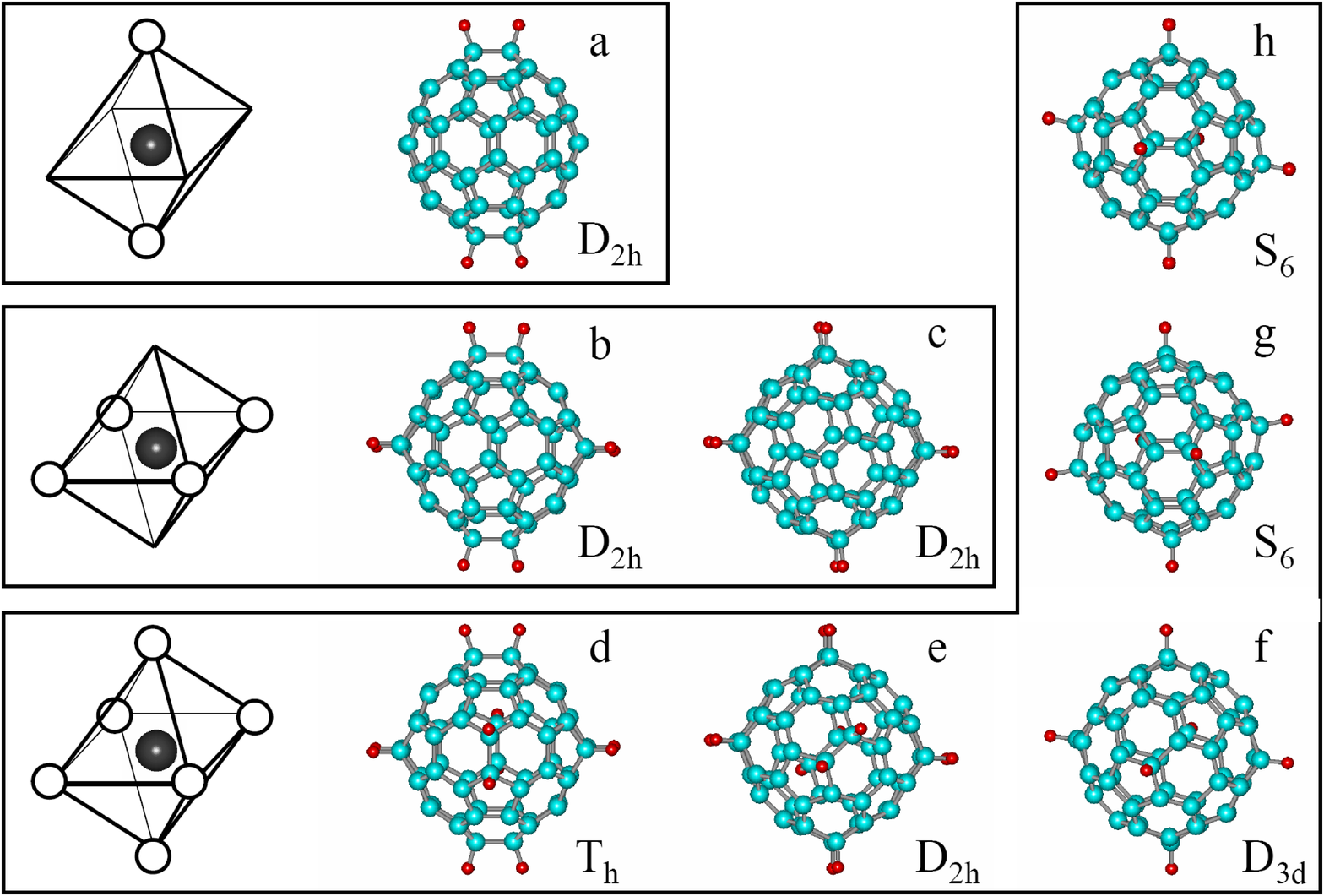}
\caption{Possible connection patterns of the fullerene in \ccp. The first column shows the arrangement of \cub\ molecules
(white spheres) which connect to a \C60\ ball (grey sphere). In the
next columns, the carbon atoms of fullerene origin are colored blue,
those of cubane origin by red. We assumed in this scheme that the
connection is four-centered, including two atoms of cubane origin.
The point group of the fullerene unit is indicated.}\label{fig:fc}
\end{figure*}

To draw the possible structures with the appropriate point groups we
recall our assumption based on structural
data\cite{pekker05,iwasiewicz07} that the local arrangement of the
molecules does not change significantly on polymerization; thus the
fullerenes must still be surrounded octahedrally by cubanes. In
addition, on polymerization the inversion center of the \C60\
molecule can be retained only if it is connected to an even number
of C$_8$H$_8$ molecules. The connection patterns selected by this
condition from the set of possible point groups are depicted in
Fig.~\ref{fig:fc}. This subset contains $T_h, S_6$, $D_{3d}$ and
$D_{2h}$.

Three types of fullerene-\cub\ connections appear in the possible
structures. In the first case (pattern $a, b$ and $d$ in the second
column of Fig.~\ref{fig:fc}) the \cub-fullerene connection involves
two adjacent carbon atoms on the double bond of the \C60\ molecule
connecting two hexagons, just as in the case of the high-pressure
high-temperature (HPHT) \C60\ polymers.\cite{nr95} The difference is
that while in those polymers a cyclobutane ring is formed on
polymerization, here  both a four-center (cyclobutane) and a
three-center (cyclopropane) ring is possible. The
second type of fullerene-\cub\ connection (pattern $c$ and $e$ in
the third column of Fig.~\ref{fig:fc}.) is formed again by two atoms
of \C60, but these lie on pentagon-hexagon bonds. It has been shown
that such a connection pattern can only exist if the ball is
opened.\cite{Schick98} As an opening was excluded based on IR
results, pattern $c$ and $e$ can be eliminated. The last type of
connection between a fullerene and a \cub\ is a single bond (pattern
$f, g$ and $h$ in the fourth column of Fig.~\ref{fig:fc}).

Next we subject these remaining structures to closer scrutiny.
Pattern $a$ was observed in the linear orthorhombic \C60\ polymer,
and $b$ in the two-dimensional tetragonal polymer.\cite{nr95} In
these polymers and in the \C60\ dimer an empirical relation holds
between the shift of the \Ag(2) mode and the number of bonds on a
single \C60\ ball: the shift is 5~\cm-1\ for every cycloaddition
connection (i.e. two adjacent single bonds).\cite{wagberg99} The
softening occurs because the bonds formed in the polymerization
reaction originate from the $\pi$-bonds of the fullerene. The shift
of 10~\cm-1\ in \ccp\ fits perfectly to pattern $a$. As
the half width of the measured peak is 7~\cm-1, it is highly
unlikely that pattern $b$ or pristine \C60 are present in \ccp.

We can rule out that each fullerene is connected to six
cubanes. In this case, because of the stoichiometry, the C$_8$H$_8$
molecule should also show sixfold coordination, which would lead to
a steric tension with six of the eight C atoms of the
hydrocarbon bound to a \C60\ molecule. Therefore structures $d, f,
g$ and $h$ would automatically imply structure $a$ to be present as
well.

According to our knowledge no fullerene compounds with the
connection pattern $d, f, g$ and $h$ have been thoroughly
investigated by vibrational spectroscopy so far. A similar well
known structure only appears in the case of pattern $d$: the
two-dimensional rhombohedral \C60\ polymer\cite{nr95} has six pairs
of $\sigma$-bonds on hexagon-hexagon bonds of the \C60\ molecule,
although arranged in a different way. The rhombohedral polymer shows
the \Ag(2) peak at 1406~\cm-1 (Ref.\citenum{davydov00}). We can
expect a shift of similar magnitude in the case of pattern $d$, but
a peak with such a shift was not observed. Another argument which
confirms the absence of pattern $d$ comes from the polarization
dependence of the Raman spectrum. If \ccp\ contained only
fullerenes with $T_h$ symmetry, then the spectrum should
show totally polarized modes, which is not the case. If, on the other hand, it
contained fullerenes with different connection patterns and
pattern $d$ were one of these, then the peaks should shift or at
least change their shape as we change the polarization. As this was
not observed either, we can again come to the conclusion that
pattern $d$ is not present in \ccp.

Up to this point we derived that \ccp\ definitely  contains
fullerene units with connection pattern $a$, but the possibility of patterns $f$, $g$,
and/or $h$ cannot be unambigously excluded. If more connection patterns
are present, then many newly activated modes should appear, which
would lead to a very rich spectrum, like e.g. that of the \C60\
photopolymer.\cite{onoe96} This is in contradiction to the observed
spectra. The presence of sixfold, besides twofold,  coordinated \C60\ would also
mean that in the frequency region of the \Ag, \Hg\ and \T1u\ modes we would have
to see at least 2, 8 and 5 modes,
respectively. Instead, we only see somewhat broader peaks as usual.
The only remaining possibility would be that all of the Raman and
infrared modes of the sixfold coordinated \C60\ units behave in a
very similar way to those of the units with pattern $a$, which would
lead to unobservable splitting. This is very unlikely since the
fullerene-\cub\ bonds in the two cases are different. Thus, based on
our infrared and Raman measurements we propose that \ccp\
consists of \cub\ molecules and fullerene molecules
connected according to pattern $a$.

The twofold coordination of the fullerene unit means that the \cub\
unit also has a coordination number of two leading to a structure
consisting of chains. We cannot derive a definite assignment as to the structure
of the cubane isomer connecting two fullerenes. One possible product, dihydropentalene,
would lead to linear chains, but there are possibilities to introduce a 90$^{\circ}$
turn as well. The simultaneous
appearance of the two would introduce disorder in all directions,
leading to the cubic and amorphous crystal structure in
accordance with x-ray diffraction.\cite{pekker05} The variety in the connecting cubane
isomers would also explain the broadening of the vibrational lines.

We can also relate the above conclusions to the structural data on
\ccm\ polymerized at various temperatures and
pressures.\cite{iwasiewicz07} Iwasiewicz-Wabnig \textit{et al.}
found two different polymer structures depending on polymerization
temperature and pressure: a pseudo-cubic and a pseudo-orthorhombic
one. They concluded from Raman spectroscopy that the two do not
differ significantly on the molecular level, but the
pseudo-orthorhombic form is more ordered since its formation
occurs at pressures where the rotation of the fullerene  balls is
sterically hindered. This leads us to believe that the D$_{2h}$
symmetry, compatible with the orthorhombic crystal structure, is
intrinsic to the polymer, and the pseudo-cubic allotrope results
from a disordered arrangement of these molecular units.

\subsection{Photochemical reaction in \ccm}

\begin{figure}[p]
\includegraphics{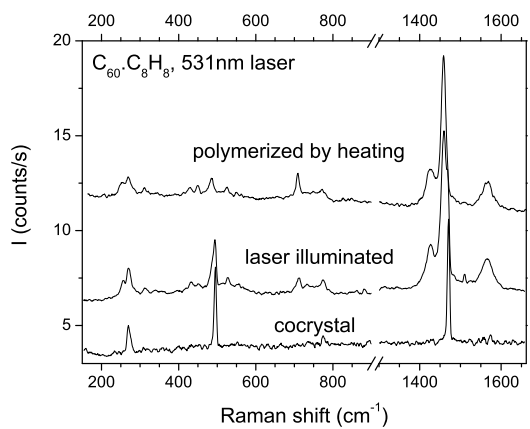}
\caption{The Raman spectrum of \ccp\ after photochemical reaction
compared to the spectrum of the cocrystal and the spectrum of
the copolymer obtained by annealing.} \label{fig:cubfp}
\end{figure}

We observed a reaction between the constituents on illumination at
room temperature similar to that taking place on heating. After
already 100~s of laser illumination in the Raman microscope
at both 531~nm and 468~nm, the intensity of the Raman peak
at 1469~\cm-1\ decreases and a new line at 1459 \cm-1\ appears. The
Raman spectrum obtained after about an hour of illumination by the
531~nm laser is depicted in Fig~\ref{fig:cubfp}. The new features in
the spectrum coincide with those of the polymer produced by
annealing. However, as we will see later, the polymerization here is
not triggered by laser-induced heating. Unfortunately we do not
observe any cubane vibrations when exciting with the laser lines at
531~nm and 468~nm, so we do not know whether cubane isomerizes the
same way as in the thermal polymerization process; we can only
deduce that the connection pattern of the fullerene is identical.

\begin{figure}[p]
\includegraphics{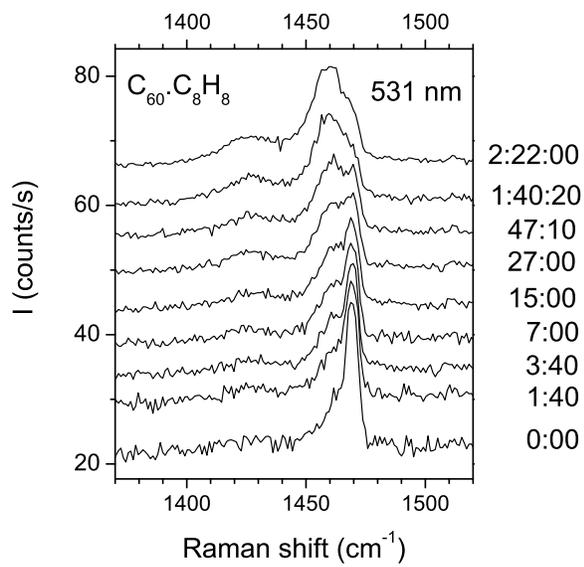}
\caption{The change of the Raman spectrum of \ccm\ on
illumination by the 531~nm laser. The time (in
hours:minutes:seconds) of the illumination is indicated on the right
hand side. } \label{fig:polproc}
\end{figure}

\begin{figure}[p]
\includegraphics{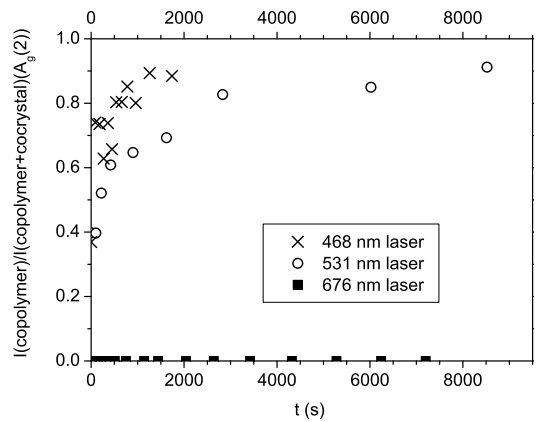}
\caption{The fractional intensity of
the \ccp\ $A_g$(2) peak as a function of illumination time for
three different lasers.} \label{fig:colfp}
\end{figure}

The gradual evolution of the new spectral pattern around the
$A_g$(2) mode during illumination is illustrated in
Fig.~\ref{fig:polproc}. We fitted the spectra with three
Lorentzians: one for the \Ag(2) mode of the monomer, one for the
\Ag(2) mode of the polymer and one for the \Hg(7) mode of the
polymer. From the obtained integrated intensity values the intensity
of the polymer \Ag(2) peak normalized to the total intensity of the
two \Ag(2) peaks was calculated. We repeated the procedure for three
exciting laser wavelengths: 531~nm, 468~nm and 676~nm (see
Fig.~\ref{fig:colfp}). We found that longer-wavelength laser lines
(676~nm or 785~nm) did not induce the reaction, therefore the effect
of laser heating can be excluded. The wavelength dependence is
analogous to that in \C60, where photopolymerization takes place on
illumination.\cite{raosci} Based on these analogies, we classify the
reaction as photo-copolymerization with excitation of C$_{60}$ as
the first step. (We note that the photochemical reaction is also the
reason why the accumulation time for the spectrum of the \ccm\
cocrystal taken at 531~nm (Fig. \ref{fig:C60Ram}) had to be shorter
than for that taken at 785~nm (Fig. \ref{fig:C60monRam}), which
accounts for the poorer statistics of the former spectrum.)

\subsection{Poly(C$_{70}$C$_8$H$_8$)}

In \hcm\ a similar irreversible change as in \ccm\ takes place on
heating to 470~K. We show the Raman and IR spectra of the reaction
product in Figs. \ref{fig:C70Ram} and \ref{fig:cc70heating} and list
the center frequencies of the peaks in Table \ref{table:c70r} and
\ref{table:c70ir}, along with the assignments of C$_{70}$ modes by
Stratmann \emph{et al}.\cite{stratman98} The reaction leads to the
disappearance of the cubane peaks from both the IR and Raman
spectra, and a new peak appears at 2946 \cm-1 in the IR spectrum. At
the same time the IR lines of the fullerene split, but the splitting
is much less than in the \C60 analogue. The Raman lines only
broaden, probably due to unresolved splitting.

We found that below 800~\cm-1 the splitting is twofold in the case
of doubly degenerate E$'_1$ modes. Above 800~\cm-1 no clear
splitting can be seen, but the lines become somewhat smeared out.
From the apparent twofold splitting of the low frequency E$'_1$
modes the loss of the fivefold axis can be concluded,
corresponding to the point group of \ch\ being $C_{2v}$ or one of
its subgroups.

The changes in the IR spectra of \hcm\ on annealing reveal a
reaction in which the cubane structure changes completely. The
resulting hydrocarbon bonds to \ch, whose cage distorts, but
remains intact. As the reaction product is insoluble in common
solvents,\cite{pekker05} it must indeed be a polymer. At this
stage of the research we cannot say anything more about the
structure of this polymer, which is partly due to the scarcity of
sound spectroscopic results on \ch\ derivatives and partly due to
the more complicated structure of \ch.

\section{Conclusions}
\label{conclusions}

The IR and Raman spectra of \ccm\ and \hcm\ were measured both in
their rotor-stator and in their polymer phases. The rotor-stator
nature of the cocrystals directly manifests itself in the
spectra being simple superpositions of those of the constituents.
Hence, van der Waals forces are the exclusive interaction between
the static cubane and rotating fullerene molecules. The slightly
lower frequency of the cubane lines can be explained on the basis of
the highly polarizable environment of the cubane molecules in these
structures.

In the IR and Raman spectra of the polymer phases the fullerene
lines are split and new lines appear, corresponding to a symmetry
lowering of the fullerene molecules whilst their cage remains
intact. As the cubane lines change dramatically during the
polymerization, we conclude that the cubane isomerizes to another
constitutional isomer, which binds to the fullerenes. According to
the vibrational spectra no \C60-\C60\ bonding occurs. The comparison
of structural and spectroscopic results allows us to identify linear
chains connected via the apical cubane as the most probable
polymerization pattern in \ccp, with possibly another
cubane isomer introducing occasional 90$^{\circ}$ turns in the chains.

Finally, we found a photochemical reaction in \ccm\ under
illumination with green or blue light. The symmetry of the fullerene
molecules in the product turns out to be the same as that in the
thermopolymer.

\section{Acknowledgments}

We gratefully acknowledge valuable discussions with G. Oszl\'anyi
and G. Bortel about x-ray diffraction measurements. This work was
supported by the Hungarian National Research Fund under Grant Nos.
OTKA T 049338 and T046700, and by the Alexander-von-Humboldt
Foundation through the Research Partnership Program 3 - Fokoop -
DEU/1009755.

\begin{table}[p]
\caption{Raman  frequencies of the \ccm\ cocrystal and \ccp\ copolymer, and
assignment\cite{bethune91,della79} of the cocrystal peaks. C stands
for cubane and F for fullerene peaks.} \label{table:c60r}
\begin{tabular}{|cc|c|}
\hline \hline \multicolumn{2}{|c|}{\ccm}&{\ccp}\\
$\nu$* (\cm-1) & assignment & $\nu$* (\cm-1) \\ \hline
 271 & F, \Hg(1) & 255 \\
 & & 272 \\
 & & 314 \\
 428 & F, \Hg(2) & 429 \\
 & & 451 \\
 495 & F, \Ag(1)& 486 \\
 & & 524 \\
 & & 560 \\
 708 & F, \Hg(3) & 711 \\
 & & 732 \\
 & & 752 \\
 770 & F, \Hg(4)& 774 \\
 904 & C, $E_g$& \\
 1000 & C, $A_{1g}$ & \\
 1072 & C, $E_g$ & \\
 1099 & F, \Hg(5)& \\
 1248 & F, \Hg(6) & \\
 1423 & F, \Hg(7) & 1426\\
 1469 & F, \Ag(2) & 1459\\
 1576 & F, \Hg(8) & 1566\\
 3008 & C, $A_{1g}$ & \\
 \hline \hline
\end{tabular}
\end{table}

\begin{table}[p]
\caption{Infrared frequencies of the \ccm\ cocrystal and \ccp\ copolymer, and
assignment\cite{bethune91,della79} of the cocrystal peaks. C stands
for cubane and F for fullerene peaks.} \label{table:c60ir}
\begin{tabular}{|cc|c|}
\hline \hline
\multicolumn{2}{|c|}{\ccm}&{\ccp}\\
$\nu$* (\cm-1) & assignment & $\nu$* (\cm-1) \\ \hline
 527 & F, \T1u(1) & 526 \\
 & & 551 \\
 & & 561 \\
 577 & F, \T1u(2) & 574 \\
 & & 705 \\
 & & 723 \\
 & & 742 \\
 & & 768 \\
 857 & C, \T1u & \\
 1181 & F, \T1u(3) & 1181 \\
 1224 & C, \T1u & \\
 1428 & F, \T1u(4)& 1424  \\
 & & 1458 \\
 2976 & C, \T1u & 2948\\
 \hline \hline
\end{tabular}
\end{table}

\begin{table}[p]
\caption{Raman  frequencies of the \hcm\ cocrystal and their
assignment according to Ref. \citenum{sun02}. All peaks are
fullerene peaks. The peaks of \hcp\ have essentially the
same center frequencies. }\label{table:c70r}
\begin{tabular}{|cc|}
\hline \hline $\nu$* (\cm-1) & assignment\cite{sun02} \\ \hline
 259 &  A$'_1$ \\
 397& A$'_1$\\
 411& E$''_1$\\
 454 &  A$'_1$ \\
 507& E$'_2$\\
 568 &  A$'_1$\\
 701& A$'_1$\\
 713& E$''_1$\\
 737 &  E$''_1$ \\
 769& E$'_2$\\
 1060& A$'_1$\\
 1182 & A$'_1$\\
 1227 & A$'_1$\\
 1256 & E$'_2$ \\
 1313 & E$''_1$ \\
 1333 & E$'_2$\\
 1368 & E$''_1$ \\
 1433 & E$''_1$ \\
 1445 & A$'_1$ \\
 1466 & A$'_1$ \\
 1512 & E$''_1$ \\
 1564 & A$'_1$ \\
 \hline \hline
\end{tabular}
\end{table}

\begin{table}[p]
\caption{Infrared frequencies of the \hcm\ cocrystal and \hcp, and
the assignment of the former according to Ref.\citenum{stratman98}.
C stands for cubane peaks, F for fullerene
peaks.}\label{table:c70ir}
\begin{tabular}{|cc|c|}
\hline \hline \multicolumn{2}{|c|}{\hcm}&{\hcp}\\
$\nu$* (\cm-1) & assignment\cite{stratman98} & $\nu$* (\cm-1) \\
\hline
 535 & F, E$'_1$ & 533 \\
  & & 541 \\
 565 & F, A$''_2$& 565 \\
  & & 569 \\
 578 & F, E$'_1$ & 578 \\
  & & 582 \\
 642 & F, E$'_1$ & 641 \\
  & & 647 \\
 674 & F, E$'_1$ & 671 \\
  & & 676 \\
  & & 763 \\
 795 & F, E$'_1$ & 776 \\
  & & 794 \\
 856 & C, \T1u & \\
 1085 & F, E$'_1$& 1086 \\
 1133 & F, A$''_2$& 1132 \\
  & & 1154 \\
  & & 1190 \\
  & & 1202 \\
  & & 1217 \\
 1202 & F, A$''_2$ & \\
 1222 & C, \T1u &  \\
 1291 & F, E$'_1$& \\
 1319 & F, A$''_2$&   \\
 1413 & F, E$'_1$& 1413 \\
 1429 & F, E$'_1$ & 1427 \\
 2974 & C, \T1u & 2964 \\
 \hline \hline
\end{tabular}
\end{table}

\begin{table}[p]
\caption{Correlation tables for the \Ag, \Hg, and \T1u\
representations of $I_h$, for the subgroups of $I_h$ containing
inversion. R denotes Raman, IR infrared active
modes.}\label{table:correl}
\begin{tabular}{|c|ccc|}
\hline \hline $I_h$ & \Ag(R) & \Hg(R) & \T1u(IR) \\
\hline  $T_h$ & \Ag(R) & $T_g$(R) + $E_g$(R) & $T_u$(IR) \\
 $S_6$ & \Ag(R) & $A_g$(R) + 2 $E_g$(R) & $A_u$(IR) + $E_u$(IR) \\
 $D_{5d}$ & $A_{1g}$(R) & $A_{1g}$(R) + $E_{1g}$(R) + $E_{2g}$(R) & $A_{2u}$(IR) + $E_{1u}$(IR) \\
 $D_{3d}$ & $A_{1g}$(R) & $A_{1g}$(R) + 2 $E_{g}$(R) & $A_{2u}$(IR) + $E_{u}$(IR) \\
 $D_{2h}$& \Ag(R) & 2$A_g$(R)+ & $B_{1u}$(IR)+$B_{2u}$(IR)+$B_{3u}$(IR) \\
 &  & +$B_{1g}$(R)+$B_{2g}$(R)+$B_{3g}$(R) &  \\
 $C_{2h}$ & \Ag(R) & 3$A_g$(R)+2$B_{g}$(R) & $A_{u}$(IR)+2$B_{u}$(IR) \\
 $C_{i}$ & \Ag(R) & 5$A_g$(R) & 3$A_{u}$(IR) \\
  \hline \hline
\end{tabular}
\end{table}
\clearpage
\clearpage
\clearpage
\clearpage
\clearpage
\clearpage
\clearpage
\clearpage
\clearpage
\end{document}